\documentclass[reprint,superscriptaddress,preprintnumbers,amsmath,amssymb,amsfonts,aps,prl,floatfix]{revtex4-2}

\usepackage{graphicx}
\usepackage[justification=RaggedRight]{caption}
\usepackage[normalem]{ulem}

\allowdisplaybreaks

\newcommand\varpm{\mathbin{\vcenter{\hbox{%
  \oalign{\hfil$\scriptstyle+$\hfil\cr
          \noalign{\kern-.3ex}
          $\scriptscriptstyle({-})$\cr}%
}}}}

\def\beq{\begin{equation}}
\def\eeq{\end{equation}}
\def\bea{\begin{eqnarray}}
\def\eea{\end{eqnarray}}
\def\nn{\nonumber}

\def\bsll{b\to s\ell^+\ell^-}
\def\bctaunu{b\to c\tau^-{\bar\nu}_\tau}
\def\btopik{B\to\pi K}
\def\ok{\overline{K}^0}

\def\tT{{\widetilde{T}}}
\def\tC{{\widetilde{C}}}
\def\tA{{\widetilde{A}}}
\def\tP{{\widetilde{P}}}

\allowdisplaybreaks

\begin{document}

\title{\boldmath Anomalies in Hadronic $B$ Decays}

\author{Rapha\"el Berthiaume}
\email{raphael.berthiaume@umontreal.ca}
\affiliation{Physique des Particules, Universit\'e de Montr\'eal, 1375 Avenue Th\'er\`ese-Lavoie-Roux, Montr\'eal, QC, Canada  H2V 0B3}

\author{Bhubanjyoti Bhattacharya}
\email{bbhattach@ltu.edu}
\affiliation{Department of Natural Sciences, Lawrence Technological University, Southfield, MI 48075, USA}

\author{Rida Boumris}
\email{rida.boumris@umontreal.ca}
\affiliation{Physique des Particules, Universit\'e de Montr\'eal, 1375 Avenue Th\'er\`ese-Lavoie-Roux, Montr\'eal, QC, Canada  H2V 0B3}

\author{Alexandre Jean}
\email{alexandre.jean.1@umontreal.ca}
\affiliation{Physique des Particules, Universit\'e de Montr\'eal, 1375 Avenue Th\'er\`ese-Lavoie-Roux, Montr\'eal, QC, Canada  H2V 0B3}

\author{Suman Kumbhakar}
\email{suman.kumbhakar@umontreal.ca}
\affiliation{Physique des Particules, Universit\'e de Montr\'eal, 1375 Avenue Th\'er\`ese-Lavoie-Roux, Montr\'eal, QC, Canada  H2V 0B3}

\author{David London}
\email{london@lps.umontreal.ca}
\affiliation{Physique des Particules, Universit\'e de Montr\'eal, 1375 Avenue Th\'er\`ese-Lavoie-Roux, Montr\'eal, QC, Canada  H2V 0B3}

\preprint{UdeM-GPP-TH-23-299}

\begin{abstract}
    In this paper, we perform fits to $B \to PP$ decays, where $B = \{B^0, B^+, B_s^0\}$ and the pseudoscalar $P = \{\pi, K\}$, under the assumption of flavor SU(3) symmetry [SU(3)$_F$]. Although the fits to $\Delta S=0$ or $\Delta S=1$ decays individually are good, the combined fit is very poor: there is a $3.6\sigma$ disagreement with the SU(3)$_F$ limit of the standard model (SM$_{\rm{SU(3)}_F}$). One can remove this discrepancy by adding SU(3)$_F$-breaking effects, but 1000\% SU(3)$_F$ breaking is required. The above results are rigorous, group-theoretically -- no dynamical assumptions have been made. When one adds an assumption motivated by QCD factorization, the discrepancy with the SM$_{\rm{SU(3)}_F}$ grows to $4.4\sigma$.
\end{abstract}

\maketitle

For the past 10$+$ years, there has been an enormous amount of interest
in the semileptonic $B$ anomalies involving the decays $\bsll$ 
($\ell = \mu, e$) and $\bctaunu$. 
Interestingly, there have also been hadronic $B$ anomalies, but these have generally 
flown under the radar. The $\btopik$ puzzle has been around for about 20 years (see Refs.~\cite{Beaudry:2017gtw,Bhattacharya:2021shk} and references therein), but discrepancies in other sets of  
hadronic decays have recently been pointed out. These include the U-spin 
puzzle \cite{Bhattacharya:2022akr}, three puzzles involving $B_s^0\to K^0 
\ok$ \cite{Amhis:2022hpm}, and a puzzle in $B_{d,s}^0 \to K^{(*)0} \overline{ 
K}^{(*)0}$ decay \cite{Biswas:2023pyw}. Of these four, the first three 
involve only $B\to PP$ decays, where $B = \{B^0, B^+, B_s^0\}$, and the 
pseudoscalar $P = \{ \pi, K \}$.  This class of $B$ decays is the focus 
of our study.

In all of these puzzles, one has a set of $B$ decays whose amplitudes are related, either by a symmetry, or simply by having the same quark-level decay. The $\btopik$ puzzle involves the four decays $B^+\to\pi^0 K^+$, $B^+\to\pi^+ K^0$, $B^0\to\pi^- K^+$, and $B^0\to\pi^0 K^0$, whose amplitudes form an isospin quadrilateral. U spin
relates the decays $B_{d,s}^0\to P^\pm P^{\prime\mp}$, where $P$ and $P'$ are each $\pi$ or $K$. And $B_s^0\to K^0\ok$ is related to
$B_s^0\to K^+ K^-$ by isospin, to $B^0\to K^0\ok$ by U spin, and to $B^+\to\pi^+ K^0$ by virtue of having the same quark-level decay. In each set of related decays, the puzzle arises because it is
found that the measured values of the observables of all the related decays are not consistent with one another.

The key point here is that all of these $B \to PP$ decays are related
to one another by flavor SU(3) symmetry [SU(3)$_F$]. By performing a
global fit to all the $B \to PP$ observables under the assumption of
SU(3)$_F$, these puzzles can be combined, and one can quantify just
how well (or poorly) the data is explained by the SU(3)$_F$ limit of the standard model (SM$_{\rm{SU(3)}_F}$).
Analyses of this type were done many years ago \cite{Chiang:2004nm},
using diagrams as the theoretical parameters and making dynamical
assumptions in order to neglect certain diagrams. But today there is
enough data that no approximations are necessary -- a full SU(3)$_F$
fit can be performed. There are even enough observables in the fit to
quantify a number of SU(3)$_F$-breaking effects. As we will see, there
are serious discrepancies with the SM.

We are interested in charmless $B \to PP$ decays, which are associated
with the transitions ${\bar b} \to {\bar u} u {\bar q}$ and ${\bar b}
\to {\bar q}$, $q = d, s$. The weak Hamiltonian is \cite{Buchalla:1995vs}
\bea
H_W &=& \frac{G_F}{\sqrt{2}} \sum_{q=d,s}\left(\lambda_u^{(q)}\left[c_1({\bar b}u)_{V-A}({\bar u}q)_{V-A}\right.\right. \nn \\
&&\hspace{-5truemm} \left.\left. +~c_2 ({\bar b}q)_{V-A}({\bar u}u)_{V-A}\right] - \lambda_t^{(q)} \sum_{i=3}^{7} c_i Q_i^{(q)} \right) ~, \label{eq:HW}
\eea
where $\lambda_{q'}^{(q)} = V_{q'b}^* V_{q'q}$, $q=d,s$, $q' =
u,c,t$. Here the $c_i$ ($i = 1$-10) are Wilson coefficients, and $Q_i^{(q)}$ represent penguin operators of two kinds: gluonic ($i = 3$-6) and electroweak ($i = 7$-10). $H_W$ transforms as a
${\bf 3^*_1}$, ${\bf 3^*_2}$, ${\bf 6}$, or ${\bf 15^*}$ of
SU(3)$_F$. The initial $B$ is a ${\bf 3}$ and the final state is
$({\bf 8} \times {\bf 8})_s = {\bf 1} + {\bf
  8} + {\bf 27}$. Putting these all together, charmless $B \to
PP$ decays are described by seven reduced matrix elements (RMEs). These are:
\bea
\lambda_u^{(q)} &:& A_1 = \langle {\bf 1} || {\bf 3^*_1} || {\bf 3} \rangle ~,~
A_8 = \langle {\bf 8} || {\bf 3^*_1} || {\bf 3} \rangle ~,~ \nn \\
\lambda_t^{(q)} &:& B_1 = \langle {\bf 1} || {\bf 3^*_2} || {\bf 3} \rangle ~,~
B_8 = \langle {\bf 8} || {\bf 3^*_2} || {\bf 3} \rangle ~, \nn \\
\lambda_u^{(q)}~{\&}~\lambda_t^{(q)} &:& R_8 = \langle {\bf 8} || {\bf 6} || {\bf 3} \rangle ~,~ 
P_8 = \langle {\bf 8} || {\bf 15^*} || {\bf 3} \rangle ~,~ \nn \\
&& ~~~~~~~~ P_{27} = \langle {\bf 27} || {\bf 15^*} || {\bf 3} \rangle~. ~~
\eea
If SU(3)$_F$ is unbroken, these RMEs are the same for $\Delta S=0$ and
$\Delta S=1$ decays. However, they can be different if
SU(3)$_F$-breaking effects are allowed.

The idea is then to express the amplitudes for all charmless $B \to
PP$ decays in terms of these seven RMEs, and then to perform a fit.
However, before doing this, we note that an equivalent description of
these SU(3)$_F$ amplitudes is provided by quark diagrams
\cite{Gronau:1994rj, Gronau:1995hn}. There are eight topologies,
representing tree ($T$), color-suppressed tree ($C$), annihilation
($A$), $W$-exchange ($E$), penguin ($P$), penguin-annihilation ($PA$),
electroweak penguin ($P_{EW}$) and color-suppressed electroweak
penguin ($P^C_{EW}$) amplitudes. $T$, $C$, $E$ and $A$ are associated
with $\lambda_u^{(q)}$, while $P_{EW}$ and $P^C_{EW}$ are associated
with $\lambda_t^{(q)}$. $P$ and $PA$ each have three pieces, related
to the flavor of the up-type quark in the loop. When CKM unitarity is
imposed to remove the $c$-quark pieces, $P_{uc}$ and $PA_{uc}$ are
associated with $\lambda_u^{(q)}$, $P_{tc}$ and $PA_{tc}$ with
$\lambda_t^{(q)}$.  Previously, it was often customary to absorb
magnitudes of CKM matrix elements into these diagrams. However, in 
this paper the CKM factors are kept separate.

In order to find how RMEs are related to diagrams, one has to compare
the expressions for amplitudes in terms of diagrams with those in
terms of RMEs \cite{Zeppenfeld:1980ex}. The five RMEs associated with
$\lambda_u^{(q)}$ are related to the six diagrams $T$, $C$ $A$, $E$,
$P_{uc}$ and $PA_{uc}$ (e.g., see Ref.~\cite{Gronau:1994rj}). These
diagrams only appear in five combinations, and it is convenient to
eliminate $E$ by defining five effective diagrams:
\bea
&&\tT \equiv T+E ~,~~ \tC \equiv C-E ~,~~ \tA \equiv A+E ~, \nn\\ 
&&\hspace{6truemm}\tP_{uc} \equiv P_{uc}-E ~,~~ {\widetilde{PA}}_{uc} \equiv PA_{uc} + E ~.
\label{effectivediagrams}
\eea
The relations between the RMEs and these effective diagrams are as
follows \footnote{We note that $A_1$, $R_8$ and $P_8$ have the opposite
  sign in Ref.~\cite{Gronau:1994rj}. This is simply a different
  convention and has no physical importance.}:
\bea
A_1    &=& \frac{1}{2\sqrt{3}}\left(-3\tT+\tC-8\tP_{uc}-12{\widetilde{PA}}_{uc}\right) ~, \nn\\
A_8    &=& \frac{1}{8}\sqrt{\frac{5}{3}}\left(-3\tT+\tC-8\tP_{uc}-3\tA\right) ~, \nn\\
R_8    &=& \frac{\sqrt{5}}{4}\left(\tT-\tC-\tA\right) ~, \nn\\
P_8    &=& \frac{1}{8\sqrt{3}}\left(\tT+\tC+5\tA\right) ~, \nn\\
P_{27} &=& - \frac{1}{2\sqrt{3}}\left(\tT+\tC\right) ~.
\eea
The relations between diagrams and the two RMEs associated with
$\lambda_t^{(q)}$ are
\beq
B_1 = -\frac{4}{\sqrt{3}}\left(\frac32 PA_{tc}+P_{tc}\right) ~,~~
B_8 = -\sqrt{\frac{5}{3}} P_{tc} ~.
\eeq

Finally, the electroweak penguin diagrams $P_{EW}$ and $P^C_{EW}$ are
also related to RMEs. But since there are only seven RMEs, and since
all of these are related to other diagrams (see above), $P_{EW}$ and
$P^C_{EW}$ must be related to these other diagrams. These EWP-tree relations, which hold in the SU(3)$_F$ limit, 
are \cite{Neubert:1998pt, Neubert:1998jq, Gronau:1998fn}
\beq
P^{(C)}_{EW} = -\frac{3}{4}\left[\frac{\Sigma_9}{\Sigma_1}(\tT+\tC\varpm\tA)\varpm\frac{\Delta_9}{\Delta_1}(\tT-\tC-\tA)\right]~,
\label{EWPtree}
\eeq
where $\Sigma_1 = c_1 + c_2, \Delta_1 = c_1 - c_2, \Sigma_9 = c_9 + c_{10},$ and $\Delta_9 = c_9 - c_{10}$. Here we have kept only the contributions from $Q_9$ and $Q_{10}$ of
Eq.~(\ref{eq:HW}).  This is justified because the Wilson coefficients of
the two other electroweak penguin operators $Q_7$ and $Q_8$ are tiny
\cite{Buchalla:1995vs}.

This shows that diagrams are equivalent to RMEs. An analysis that uses
diagrams to parametrize amplitudes is therefore completly rigorous
from a group-theoretical point of view. One advantage of diagrams over
RMEs is that it is straightforward to work out the contribution of any
diagram to a given decay amplitude. It is not necessary to compute the
SU(3)$_F$ Clebsch-Gordan coefficients, which can be tricky.

Another advantage is that one can estimate the relative sizes of
different diagrams. For example, it has been argued that $E$, $A$ and
$PA$ are much smaller than the other diagrams because they involve an
interaction with the spectator quark \cite{Gronau:1994rj,
  Gronau:1995hn}, and so can (often) be neglected. But this is also
problematic: results that use dynamical assumptions such
as this are not rigorous group-theoretically. In addition, one has to
worry about whether the assumptions remain valid when rescattering
effects are included.

In this paper, we make no such assumptions. The amplitudes are
parametrized in terms of all the diagrams, and we perform fits to the
data. The sizes of the diagrams are fixed by the data. It is only at
the end that we examine the effects of adding dynamical assumptions.

\begin{table}[!htbp]
\renewcommand*{\arraystretch}{2}
\begin{center}
\begin{tabular}{|l@{}|c@{}c@{}c@{}c c@{}|c@{}c@{}c@{}c|} \hline
Decay & \multicolumn{5}{c|}{$\lambda^{(d)}_u$} & \multicolumn{4}{c|}{$\lambda^{(d)}_t$} \\ \cline{2-10}
Mode  & $\tT$ & $\tC$ & $\tP_{uc}$ & $\tA$ & ${\widetilde{PA}}_{uc}$ & ${P}_{tc}$ & ${PA}_{tc}$ & ${P_{EW}}$ & ${P_{EW}^C}$ \\ \hline\hline
$B^+\to\ok K^+$ &0 &0 &1  &1 &0 &1  &0 & 0 & $-\frac{1}{3}$\\
$B^+\to\pi^0\pi^+$ &$-\frac{1}{\sqrt{2}}$&$-\frac{1}{\sqrt{2}}$&0&0&0&0&0 & $-\frac{1}{\sqrt{2}}$ &$-\frac{1}{\sqrt{2}}$  \\ \hline
$B^0\to K^0\ok$ &0 &0 & 1 &0 &1 & 1 & 1 & 0 & $-\frac{1}{3}$ \\
$B^0\to\pi^+\pi^-$ & $-1$  &0 & $-1$ &0& $-1$ & $-1$ & $-1$  & 0 & $-\frac{2}{3}$\\
$B^0\to\pi^0\pi^0$ &0 &$-\frac{1}{\sqrt{2}}$&$\frac{1}{\sqrt{2}}$ &0 &$\frac{1}{\sqrt{2}}$ & $\frac{1}{\sqrt{2}}$ & $\frac{1}{\sqrt{2}}$ & $-\frac{1}{\sqrt{2}}$ & $-\frac{1}{3\sqrt{2}}$\\ 
$B^0\to K^+K^-$ &0&0 &0 &0& $-1$ &0 & $-1$  & 0 & 0\\ \hline
$B_s^0\to\pi^+K^-$  & $-1$ &0 & $-1$  &0 &0 & $-1$  &0 & 0 & $-\frac{2}{3}$\\
$B_s^0\to\pi^0\ok$ &0&$-\frac{1}{\sqrt{2}}$&$\frac{1}{\sqrt{2}}$&0&0&$\frac{1}{\sqrt{2}}$&0 & $-\frac{1}{\sqrt{2}}$ & $-\frac{1}{3\sqrt{2}}$\\ \hline
\end{tabular}
\end{center}
\caption{Decomposition of $\Delta S = 0$ $B \to PP$ decay amplitudes in terms of diagrams.\label{DeltaS=0amps}}
\end{table}

There are eight $B \to PP$ decays with $\Delta S=0$ and eight with
$\Delta S=1$. The decomposition of their amplitudes in terms of
diagrams is given in Tables \ref{DeltaS=0amps} and \ref{DeltaS=1amps},
respectively. Diagrams for $\Delta S=0$ and $\Delta S=1$ processes are
respectively written without and with primes. Of course, in the limit
of perfect SU(3)$_F$ symmetry, $\tT' = \tT$, etc.

\begin{table}[!htbp]
\renewcommand*{\arraystretch}{2}
\begin{center}
\begin{tabular}{|l@{}|c@{}c@{}c@{}c@{}c@{}|c@{}c@{}c@{}c|} \hline
Decay & \multicolumn{5}{c|}{$\lambda^{(s)}_u$} & \multicolumn{4}{c|}{$\lambda^{(s)}_t$}\\ \cline{2-10}
Mode  & $\tT'$ & $\tC'$ & $\tP'_{uc}$ & $\tA'$ & ${\widetilde{PA}}'_{uc}$ & $P'_{tc}$ & $PA'_{tc}$ & $P'_{EW}$ & ${P_{EW}^{\prime C}}$ \\ \hline \hline
$B^+\to\pi^+K^0$ &0 &0 &1  &1 &0 &1  &0 & 0 & $-\frac{1}{3}$ \\
$B^+\to\pi^0K^+$ &$-\frac{1}{\sqrt{2}}$&$-\frac{1}{\sqrt{2}}$&$-\frac{1}{\sqrt{2}}$&$-\frac{1}{\sqrt{2}}$&0&$-\frac{1}{\sqrt{2}}$&0& $-\frac{1}{\sqrt{2}}$ & $-\frac{\sqrt{2}}{3}$\\ \hline
$B^0\to\pi^-K^+$ & $-1$  &0 & $-1$  &0 &0 & $-1$  & 0 & 0 & $-\frac{2}{3}$ \\
$B^0\to\pi^0K^0$ &0 &$-\frac{1}{\sqrt{2}}$ &$\frac{1}{\sqrt{2}}$ &0&0& $\frac{1}{\sqrt{2}}$ &0 & $-\frac{1}{\sqrt{2}}$ & $-\frac{1}{3\sqrt{2}}$ \\ \hline
$B_s^0\to K^+K^-$ & $-1$ &0 & $-1$  &0 & $-1$  & $-1$  & $-1$ & 0 & $-\frac{2}{3}$\\
$B^0_s\to K^0\ok$&0&0 &1 &0&1&1 &1 & 0 & $-\frac{1}{3}$ \\
$B^0_s\to\pi^+\pi^-$ &0 &0 & 0 &0 & $-1$  & 0 & $-1$ & 0 & 0\\
$B^0_s\to\pi^0\pi^0$ &0&0&0&0&$\frac{1}{\sqrt{2}}$&0&$\frac{1}{\sqrt{2}}$ & 0 & 0\\ \hline
\end{tabular}
\end{center}
\caption{Decomposition of $\Delta S = 1$ $B \to PP$ decay amplitudes in terms of diagrams.\label{DeltaS=1amps}}
\end{table}

Of the 16 charmless $B \to PP$ decays, 15 have been observed. Their
measurements have given rise to a large number of observables
(CP-averaged branching ratios or ${\cal B}_{CP}$, direct CP asymmetries or $A_{CP}$, and indirect CP asymmetries or $S_{CP}$). A complete
list of these observables, along with their present experimental
values, can be found in Table \ref{observables_meas}. In terms of the theoretical parameters, the
observables are defined as
\bea
& {\cal B}_{CP} = F_{\rm PS} \, (|A|^2 + |{\bar A}|^2) ~, & \nn\\
& {\rm where} ~~ F_{\rm PS} = 
\frac{\sqrt{m_B^2 - (m_{P_1} + m_{P_2})^2}
  \sqrt{m_B^2 - (m_{P_1} - m_{P_2})^2} \, S}{32\pi \, m_B^3\, \Gamma_B} ~, & \nn\\
& A_{CP} = \frac{|{\bar A}|^2 - |A|^2} {|{\bar A}|^2 + |A|^2} ~~,~~~~
S_{CP} = 2 {\rm Im} \left(\frac{q}{p} \frac{ {\bar A} A^*}{|{\bar A}|^2 + |A|^2} \right) ~. &
\label{FPSdef}
\eea
Here $A$ and ${\bar A}$ are the amplitudes for $B \to PP$ and its
CP-conjugate process, respectively, $S$ is a statistical factor
related to identical particles in the final state, and $q/p = \exp(-2
i \phi_M)$, where $\phi_M$ is the weak phase of $B_q^0$-${\bar B}_q^0$
mixing.  Note that, for the direct CP asymmetry, some experiments
present the result for $C_{CP} = -A_{CP}$. In the Tables, we have
added the appropriate minus signs, so that all results are for
$A_{CP}$. Also, in the fits, $S_{CP}$ is multiplied by $\eta_{CP}$, the CP of the final state. In general $\eta_{CP} = 1$. The only exception is the final state $\pi^0 K_S$, for which $\eta_{CP} = -1$.

\begin{table}[!htbp]
\begin{center}
\begin{tabular}{|l|c|c|c|} \hline
Decay & ${\cal B}_{CP}$ ($\times 10^{-6}$) & $A_{CP}$ & $S_{CP}$ \\ \hline\hline
$B^+\to K^+\ok$ & 1.31$\pm$0.14& 0.04$\pm$0.14$^\dagger$ & \\
$B^+\to\pi^+\pi^0$& 5.59$\pm$0.31 & 0.008$\pm$0.035 & \\ \hline
$B^0\to K^0\ok$    & 1.21$\pm$0.16$^\dagger$ & 0.06$\pm$0.26 & $-1.08\pm$0.49\\
$B^0\to\pi^+\pi^-$ & 5.15$\pm$0.19 & 0.311$\pm$ 0.030 & $-0.666\pm$ 0.029 \\
$B^0\to\pi^0\pi^0$ & 1.55$\pm$ 0.16 & 0.30$\pm$0.20 & \\
$B^0\to K^+K^-$ & 0.080$\pm$0.015 &  &  \\ \hline
$B^0_s\to\pi^+K^-$ & $5.90^{+0.87}_{-0.76}$ & 0.225$\pm$0.012 & \\
$B^0_s\to\pi^0\ok$ & & & \\ \hline
\hline
$B^+\to\pi^+K^0$ & 23.52$\pm$0.72 & $-0.016\pm$0.015 & \\
$B^+\to\pi^0K^+$ & 13.20$\pm$0.46 & 0.029$\pm$0.012 & \\ \hline
$B^0\to\pi^-K^+$ & 19.46$\pm$0.46 & $-0.0836\pm$0.0032 & \\  $B^0\to\pi^0K^0$ & 10.06$\pm$0.43 & $-0.01\pm$0.10 & $0.57\pm$0.17 \\ \hline
$B^0_s\to K^+K^-$ & $26.6^{+3.2}_{-2.7}$ & $-0.17\pm$0.03 & 0.14$\pm$0.03 \\
$B^0_s\to K^0\ok$ & 17.4$\pm$3.1 & & \\
$B^0_s\to\pi^+\pi^-$ & $0.72^{+0.11}_{-0.10}$ & & \\
$B^0_s\to\pi^0\pi^0$ & 2.8$\pm$2.8$^*$ & & \\ \hline
\end{tabular}
\end{center}
\caption{Measured values of ${\cal B}_{CP}$, $A_{CP}$, and $S_{CP}$ in $\Delta S=0$ (upper table) and $\Delta S=1$ (lower table) $B \to PP$ decays. The $^\dagger$ indicates data taken from the Particle Data Group \cite{ParticleDataGroup:2022pth}, the $^*$ indicates data taken from Ref.~\cite{Belle:2023aau}. All other data are taken from HFLAV \cite{HFLAV:2022pwe}.
\label{observables_meas}}
\end{table}

Consider first $\Delta S=0$ decays. The amplitudes are a function of 7
diagrams, corresponding to 13 unknown theoretical parameters (7
magnitudes, 6 relative strong phases). From Table
\ref{observables_meas}, we see that there are 15 measured
observables.  The amplitudes also depend on the weak phases $\gamma$,
$\beta$ (in $B^0$-${\bar B}^0$ mixing) and $\phi_s$ (in $B_s^0$-${\bar
  B}_s^0$ mixing), as well as on the CKM matrix elements involved in
$\lambda_{u,t}^{(q)}$. Values for all of these quantities, including
errors, are taken from the Particle Data Group (PDG)
\cite{ParticleDataGroup:2022pth}.

As the quantities taken from the PDG are ``known,'' we therefore
effectively have 15 equations in 13 unknowns, so we can do a fit. The
fit is performed using the program {\tt MINUIT}
\cite{James:1975dr,*James:2004xla,*James:1994vla,*iminuit}. We find an excellent
fit: the $\chi_{\rm min}^2/{\rm d.o.f.} = 0.35/2$, for a $p$-value of
0.84. The SM$_{\rm{SU(3)}_F}$ therefore has no difficulty explaining the $\Delta S=0$
data.

Turning to $\Delta S=1$ decays, there are again 13 unknown theoretical
parameters, along with 15 measured observables (Table
\ref{observables_meas}), so a fit can be performed. Here the fit is
slightly worse, but still perfectly acceptable: $\chi_{\rm min}^2/{\rm
  d.o.f.} = 1.8/2$, for a $p$-value of 0.40.

If one assumes perfect SU(3)$_F$ symmetry, the diagrams in $\Delta
S=0$ and $\Delta S=1$ decays are the same. We can therefore perform a
fit including all the data -- we have 30 equations in 13 unknowns. But
now a serious problem arises: the best fit has $\chi_{\rm min}^2/{\rm
  d.o.f.} = 43.8/17$, for a $p$-value of $3.6 \times 10^{-4}$. This
means that the data disagrees with the SM$_{\rm{SU(3)}_F}$ at the level of
$3.6\sigma$.

We stress again that no dynamical assumptions have been made regarding the diagrams. This result is completely rigorous from a group-theoretical point of view.

Note that a similar $B\to PP$ fit including $\eta$ and $\eta'$ mesons was performed in Ref.~\cite{Huber:2021cgk} using the formalism of Ref.~\cite{He:2018php}, and a good fit was found. However, in this analysis, the diagrams $P_{EW}$ and $P^C_{EW}$ were allowed to vary freely; the EWP-tree relations [Eq.~(\ref{EWPtree})], which hold in the SU(3)$_F$ limit, were not imposed. (We confirm that, if $P_{EW}$ and $P^C_{EW}$ are left free in our fit, a good fit is found. However, the EWP-tree relations are badly broken.)

Now, our result raises an obvious question. We know that SU(3)$_F$ is broken
in the SM. What is usually quoted as evidence is the fact that
$f_K/f_\pi - 1 = ~\sim 20\%$. That is, we naively expect
SU(3)$_F$-breaking effects at this level. If such effects were
included, perhaps that would remove the discrepancy.

Fortunately, the fit contains enough information to address this
question. Above, we found that, when one considers only $\Delta S = 0$
or $\Delta S = 1$ decays, the fits are good. In Table
\ref{tab:params_diags}, for each fit we show the best-fit values of
the magnitudes of the diagrams. In the SU(3)$_F$ limit, the diagrams
in $\Delta S = 0$ decays ($D$) are the same as those in $\Delta S = 1$
decays ($D'$). Thus, the ratios $|D'/D|$ provide an indication of the
level of SU(3)$_F$ breaking required for the SM to explain the data.

These ratios are also shown in Table \ref{tab:params_diags}. For the
diagrams associated with $\lambda_u^{(q)}$, the average of the $|D'/D|$ central values is 11.7. For some of these ratios, the errors are large, so that the ratio is consistent with unity. However, these errors are also highly correlated: if one ratio is forced to be 1, another ratio will become even larger than its central value. The upshot is that at least one of the ratios (and probably more than one) is $\sim 10$.

But this corresponds to 1000\% SU(3)$_F$ breaking! This is obviously much
larger than the $\sim 20\%$ of $f_K/f_\pi$. Thus, if the SM really does explain the data, then
either the $3.5\sigma$ discrepancy is simply a
statistical fluctuation (involving several different decays), or the SM breaks flavor SU(3)$_F$ symmetry at
an unexpectedly large level. This is the anomaly in hadronic $B$ decays.
\begin{table}[!htbp]
\begin{center}
\begin{tabular}{|c|c|c|c|c|} \hline
& $|\tT|$ & $|\tC|$ & $|\tP_{uc}|$ & $|\tA|$\\ \cline{2-5}
Fit & $4.0\pm0.5$ & $6.6\pm0.7$ & $3\pm4$ & $6\pm5$ \\ \cline{2-5}
$\Delta S = 0$ & $|{\widetilde{PA}}_{uc}|$ & $|P_{tc}|$ & $|PA_{tc}|$ & \\ \cline{2-5}
& $0.7\pm0.8$ & $0.8\pm0.4$ & $0.2\pm0.4$ & \\ \hline\hline         
& $|\tT'|$ & $|\tC'|$ & $|\tP'_{uc}|$ & $|\tA'|$ \\ \cline{2-5} 
Fit & $48\pm14$ & $41\pm14$ & $48\pm15$ & $81\pm28$ \\ \cline{2-5}
$\Delta S = 1$ & $|{\widetilde{PA}}'_{uc}|$ & $|P'_{tc}|$ & $|PA'_{tc}|$ & \\ \cline{2-5}
& $7\pm4$ & $0.78\pm0.16$ & $0.24\pm0.04$ & \\ \hline\hline
& $|\tT'/\tT|$ & $|\tC'/\tC|$ & $|\tP'_{uc}/\tP_{uc}|$ & $|\tA'/\tA|$ \\ \cline{2-5}
& $12\pm4$ & $6.6\pm2.2$ & $16\pm22$& $14\pm13$ \\ \cline{2-5}
& $|{\widetilde{PA}}'_{uc}/{\widetilde{PA}}_{uc}|$ & $|P'_{tc}/P_{tc}|$ & $|PA'_{tc}/PA_{tc}|$ & \\ \cline{2-5}
& $10\pm13$ & $0.97\pm0.52$& $1.3\pm2.7$ & \\ \hline\hline
& $|\tT|$ & $|\tC|$ & $|\tP_{uc}|$ & $|\tA|$ \\ \cline{2-5} 
Fit & $4.7\pm0.5$ & $5.8\pm0.6$ & $2.1\pm0.5$ & $4.2\pm0.7$ \\ \cline{2-5}
SU(3)$_F$ & $|{\widetilde{PA}}_{uc}|$ & $|P_{tc}|$ & $|PA_{tc}|$ & \\ \cline{2-5}
& $0.70\pm0.09$ & $1.15\pm0.04$ & $0.214\pm0.018$ & \\ \hline
\end{tabular}
\end{center}
    \caption{Best-fit values of the magnitudes of the diagrams in units of keV for the
      $\Delta S=0$ and $\Delta S=1$ fits, as well as for the fit with
      unbroken SU(3)$_F$.
    \label{tab:params_diags}}
\end{table}

The large SU(3)$_F$ breaking seen in the fit is actually a reflection of large SU(3)$_F$ breaking in the experimental data. The present data give
\bea
- \frac{\delta_{\rm CP}(B^0_s\to K^+K^-)}{\delta_{\rm CP}(B^0\to \pi^+\pi^-)} &=& 2.90 \pm 0.69 ~, \nn\\
- \frac{\delta_{\rm CP}(B^0_s\to K^+K^-)}{\delta_{\rm CP}(B^0_s\to \pi^+K^-)} &=& 3.43 \pm 0.91 ~,
\label{expSU3break}
\eea
where $\delta_{\rm CP}(B_q\to PP') = A_{CP} {\cal B}_{CP}/F_{\rm PS}$ [see Eq.~(\ref{FPSdef})]. In the SU(3)$_F$ limit, both of these ratios are expected to equal 1\footnote{In fact, the second ratio is predicted to equal 1 only if the ${\widetilde{PA}}$ diagram is neglected (see Ref.~\cite{Bhattacharya:2022akr}). But since the data show that ${\widetilde{PA}}$ is much smaller than the other diagrams (see Table \ref{tab:params_diags}), this is a good approximation.} \cite{Grossman:2013lya, Gronau:2013mda}. Thus, the above experimental results each indicate $\sim 300\%$ SU(3)$_F$ breaking. Our analysis shows that, when all decays are examined simultaneously, the net SU(3)$_F$-breaking effect is quite a bit larger. It is also expected that $- \delta_{\rm CP}(B^0_s\to K^0 {\bar K}^0)/\delta_{\rm CP}(B^0\to K^0 {\bar K}^0) = 1$ in the SU(3)$_F$ limit. This ratio has not yet been measured, but according to our analysis, it should also exhibit very large SU(3)$_F$ breaking.

The sizeable width difference in the $B^0_s$ system, $\Delta\Gamma_s$, modifies the branching ratios of certain $B^0_s\to f$ decays extracted from untagged samples by a correction factor involving the $\Delta\Gamma_s$-dependent CP asymmetry, $A^f_{\Delta\Gamma}$ \cite{DeBruyn:2012wj}. However, $A^f_{\Delta\Gamma}$ has been measured in only one decay, $B^0_s\to K^+K^-$. The inclusion
of this correction factor increases ${\cal B}(B^0_s\to K^+K^-)$ by 8\%, making the SU(3)$_F$-breaking effects in Eq.~(\ref{expSU3break}) even larger and the fits considerably worse. Thus, it is likely that the discrepancy with the SM$_{SU(3)_F}$, and the size of SU(3)F breaking required to remove this discrepancy, are even larger than described above.

But this is not all. Up to now, the analysis has been completely
rigorous, group-theoretically -- no dynamical assumptions
were have been made regarding the diagrams. Returning to Table
\ref{tab:params_diags}, we see that, although the $\Delta S = 0$ and
$\Delta S = 1$ fits are good, they require values for the diagrams
that are well outside theoretical expectations.

As noted earlier, it has been argued that $E$, $A$ and $PA$ are
negligible compared to the dominant diagrams \cite{Gronau:1994rj,
  Gronau:1995hn}. For ${\widetilde{PA}}$ and ${\widetilde{PA}}'$, this
is reasonably borne out by the data: $|{\widetilde{PA}}_{uc}/\tT|$ and
$|{\widetilde{PA}}'_{uc}/\tT'|$ are both quite a bit smaller than 1. Note that, since
${\widetilde{PA}}_{uc} \equiv PA_{uc} + E$
[Eq.~(\ref{effectivediagrams})], technically $PA_{uc}$ and $E$ could
both be large. But in order to obtain the small $|{\widetilde{PA}}_{uc}|$, this would then require a fine-tuned cancellation between these two diagrams. A more natural assumption is that
$|PA_{uc}|$ and $|E|$ are both of the order of
$|{\widetilde{PA}}_{uc}|$. Furthermore, $|PA_{tc}|$ is small. The
data therefore largely confirm the theoretical expectation that $E$ and $PA$
are much smaller than the dominant diagrams. On the other hand, in
Table \ref{tab:params_diags}, we see that $|\tA/\tT|$ and
$|\tA'/\tT'|$ are both $O(1)$. This is very strange -- why would $A$
be large, while $E$ and $PA$ are small?

Another curious result is related to the ratio $|C/T|$. Naively, we
expect $|C/T| = 1/3$, simply by counting colors. This expectation
is borne out by theoretical calculations. In QCD factorization, this
ratio is computed for $B \to \pi K$ decays ($\Delta S=1$). It is found
that $|C'/T'| \simeq 0.2$ at NLO \cite{Beneke:2001ev}, while at NNLO,
$0.13 \le |C'/T'| \le 0.43$, with a central value of $|C'/T'| = 0.23$,
very near its NLO value \cite{Bell:2007tv, Bell:2009nk, Beneke:2009ek, Bell:2015koa}.  

On the other hand, the fits of Table
\ref{tab:params_diags} have $|\tC/\tT| = 1.65$ ($\Delta S = 0$),
$|\tC'/\tT'| = 0.85$ ($\Delta S = 1$), and $|\tC/\tT| = 1.23$
(SU(3)$_F$). It is true that $\tT$ and $\tC$ include contributions
from $E$, but since $E$ has been shown to be small, $|\tC/\tT| \simeq
|C/T|$, and similarly for the primed diagrams.

If we fix $|\tC^{(\prime)}/\tT^{(\prime)}|$ to 0.2 and redo the fits,
we now find that the fit of $\Delta S=1$ decays is worse than
before, but still acceptable: $\chi_{\rm min}^2/{\rm d.o.f.} = 6.8/3$,
for a $p$-value of 0.08. (But note that $\tA'$ is now the largest diagram in this fit, with $|\tA'/\tT'| = 1.6$.) On the other hand, the fit of $\Delta S=0$
decays is considerably worse: $\chi_{\rm min}^2/{\rm d.o.f.} =
18.8/3$, for a $p$-value of $3.1 \times 10^{-4}$, corresponding to a
discrepancy with the SM$_{\rm{SU(3)}_F}$ of $3.6\sigma$. Finally, if one assumes
perfect SU(3)$_F$ symmetry, the best fit has $\chi_{\rm min}^2/{\rm
  d.o.f.} = 55.8/18$, for a $p$-value of $9.4 \times 10^{-6}$. The
discrepancy with the SM$_{\rm{SU(3)}_F}$ has grown to $4.4\sigma$.

\begin{figure}[!htbp]
\begin{center}
\includegraphics[width=0.48\textwidth]{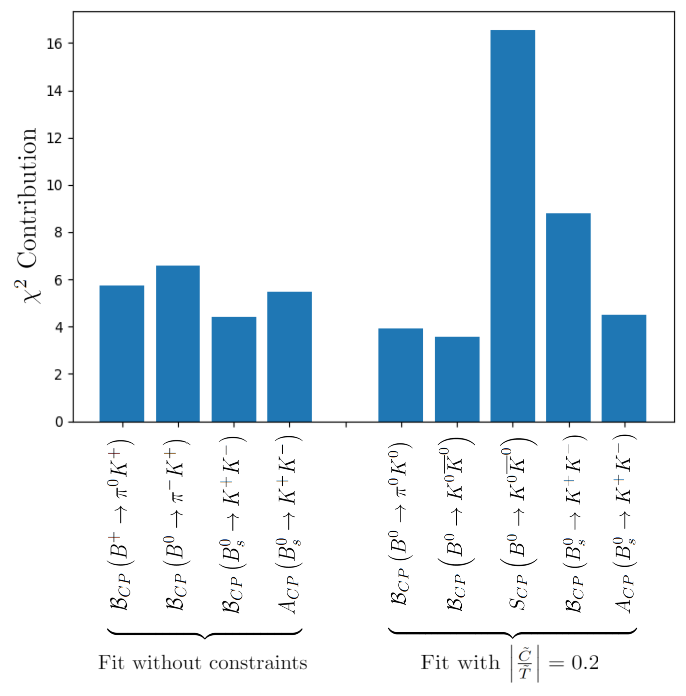}
\end{center}
\caption{Observables providing the largest $\chi^2$ contributions for the global fits with $|\tC/\tT|$ unconstrained (left) and $|\tC/\tT| = 0.2$ (right).}
\label{Observables_chi2}
\end{figure}

As we have seen, in the global fit to both $\Delta S=0$ and $\Delta S=1$ decays, the discrepancy with the SM$_{\rm{SU(3)}_F}$ is $3.6\sigma$ if $|\tC/\tT|$ is unconstrained, and it jumps to $4.4\sigma$ if $|\tC/\tT|$ is fixed to be 0.2. In Fig.~\ref{Observables_chi2}, we identify the observables that contribute the most to the $\chi^2$ of each of these fits. On the whole, the large-$\chi^2$ observables are different for the two fits; the only ones that are important for both fits are the CP-averaged branching ratio and direct CP asymmetry of $B_s^0 \to K^+ K^-$. This is unsurprising, given that this decay figures in both experimental results exhibiting large SU(3)$_F$ breaking [see Eq.~(\ref{expSU3break})].

Note that $\Delta S=1$ decays play a particularly important role in these discrepancies. This suggests that there may be new-physics contributions to $b \to s u {\bar u}$ and $b \to s d {\bar d}$. Perhaps there is a connection with the semileptonic $\bsll$ anomalies.

To sum up, assuming unbroken flavor SU(3) symmetry, a global fit to
all $B \to PP$ data finds a discrepancy with the SM$_{\rm{SU(3)}_F}$ at the level of
$3.6\sigma$. This discrepancy can be removed by allowing for
SU(3)$_F$-breaking effects, but 1000\% SU(3)$_F$ breaking is
required, i.e., parameters that are equal in the SU(3)$_F$ limit must
now differ by a factor of ten. These results are group-theoretically
rigorous -- no dynamical assumptions have been made. But
if one also requires that $|C/T| = 0.2$, which is the predicted value in QCD
factorization, the discrepancy with
the SM$_{\rm{SU(3)}_F}$ grows to $4.4\sigma$. These are the anomalies in hadronic $B$ decays. They strongly hint that new physics is present in these decays.

{\bf Acknowledgments}: We thank Marianne Bouchard for checking the fit with the $P_{EW}$ and $P^C_{EW}$ diagrams free. This work was financially supported by NSERC of
Canada (RB, RB, AJ, SK, DL) and by the National Science Foundation,
Grant No. PHY-2310627 (BB).

\bibliography{prlref}

\end{document}